# New Results from RENO

**Seon-Hee Seo[1] for the RENO Collaboration**
*Seoul National University*
*Department of Physics and Astronomy*
*1 Gwanak-ro, Gwanak-gu, Seoul, 151-747, Korea*
E-mail: `shseo@phya.snu.ac.kr`

RENO (Reactor Experiment for Neutrino Oscillation) is an experiment dedicated to measure the smallest neutrino mixing angle $\theta_{13}$ using reactor neutrinos in Korea. Our first result measured in 2012 using about 220 live days of data showed non-zero $\theta_{13}$ value with 4.9 $\sigma$ significance. In March 2013 we updated our first result with improvements in both statistical and systematic errors using 403 live days of data. The measured value using rate-only analysis is $\sin^2 2\theta_{13}$ = 0.100 +/- 0.010 (stat) +/- 0.015 (sys.) corresponding to 6.3 $\sigma$ significance. RENO has been taking data almost continuously since August 2011 and we have reached more than 800 live days of data that is currently being analyzed.

---

[1] *Speaker at the XV Workshop on Neutrino Telescopes*
*March   11-15, 2013*
*Venice, Italy*

# 1. Introduction

Neutrino oscillation is well known [1-5] and is described by Pontecorvo-Maki-Nakagawa-Sakata (PMNS) matrix [6,7] which transforms the mass eigen state (m1, m2, and m3) of neutrinos to the flavor eigen state ($\nu_e$, $\nu_\mu$, $\nu_\tau$). Since after the first mixing angle ($\theta_{23}$) measurement in 1998 [2,3], there had been only indications of non zero $\theta_{13}$ value [8-10] until recently. In 2012 the $\theta_{13}$ value was finally measured by Daya Bay [11] and RENO [12]. The remaining oscillation parameters to be measured in the PMNS matrix are the neutrino mass hierarchy (the sign of neutrino mass squared difference, i.e., $+/-\Delta m_{13}^2$) and the leptonic CP violation phase angle ($\delta_{CP}$). The relatively large value of $\theta_{13}$ implies that it is not impossible to measure the two physical quantities even though very challenging [13-18]. The precise measurements of already measured mixing angles are still important since the mixing parameters are entangled together and thus can constrain other neutrino mixing parameters as well as mass hierarchy and $\delta_{CP}$ measurements.

In reactor neutrino experiments, to avoid a problem of absolute electron anti-neutrino flux deficit [19-21] and/or the absolute flux uncertainty which is at least 5 % level [22] affecting $\theta_{13}$ measurement, it is desirable to use two identical detectors: one in near and the other in far distances from the reactors. The neutrino flux measured in a near detector will be used to predict neutrino flux without oscillation in a far detector so that the ambiguity and uncertainty of the absolute neutrino flux will be canceled out. It is worthy to note that RENO was the first reactor neutrino experiment which started taking data using both near and far detectors in August 2011.

In this paper we report our new measurement on $\sin^2 2\theta_{13}$ value using 403 live days of data.

# 2. Experimental Setup

RENO detectors are located in Hanbit reactor site in Yonggwang, about 300 km southwest from Seoul. In the Hanbit reactor site there are a total of six reactors aligned with equidistance of about 260 m and its total maximum thermal power is 16.8 $GW_{th}$. RENO near (far) detector is located under a small hill , corresponding to ~120 (450) m.w.e. (meter water equivalent) overburden, and it is 294 (1383) m away from the center of the reactor array. The distances were measured using GPS and total station methods. The errors in the distance measurements is less than 10 cm, which results in less than 1 % error in neutrino flux estimation.

# 3. The RENO Detector

The near and far detectors of RENO were designed and built to be identical to reduce the systematic error. As shown in Fig. 1 each detector consists of four coaxial layers of cylindrical vessels (with different radiuses and heights) and each cylindrical vessel contains different liquid to serve their own purposes. These layers are target, gamma-catcher, buffer, and veto from the innermost to the outermost order. In subsection 3.1 each layer is explained more in detail. In subsection 3.2 and 3.3 our energy scale calibration and detector stability are described, respectively.

### 3.1 Detector layout

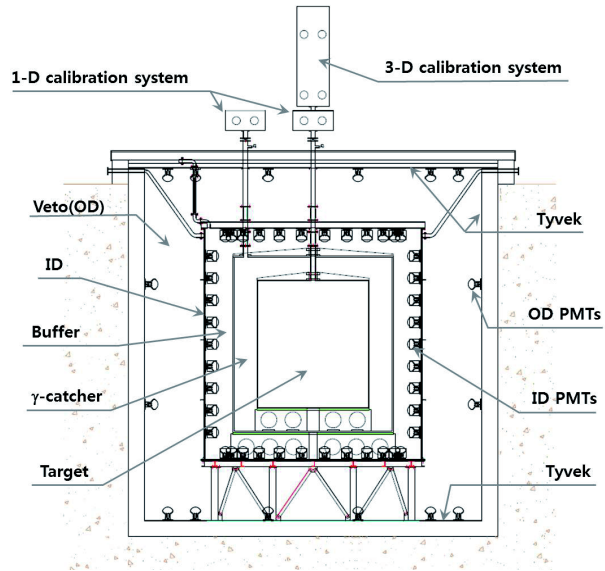

Fig. 1. A schematic view of the RENO detector layered with four cylindrical vessels filled with different liquids. See text for more details on the detector components.   There are two chimneys as passages of radioactive sources for the energy calibration.

As a result of nuclear fission from reactor cores, electron anti-neutrinos are produced about $10^{20}$ neutrinos per $GW_{th}$.   To detect these neutrinos we use the inverse beta decay (IBD) process on a proton target, which is a typical method used by reactor neutrino experiments. The target material we use is liquid scintillator (16 ton of Liquid Arkil Benzen [23]) contained in an acrylic cylindrical vessel (Radius = 1.4 m, Height = 3.2 m).   The liquid scintillator target is doped with ~0.1 % Gadolinium (Gd) to capture neutrons from the IBD processes.   The target is surrounded by gamma-catcher (Radius = 2.0 m, Height = 4.4 m) which has only liquid scintillator (30 ton) without Gd doping.   The purpose of gamma-catcher is to catch gammas from IBD process (either positron or neutron, or both) occurred outside target.   The gamma-catcher is surrounded by buffer (Radius = 2.7 m, Height = 5.8 m) which contains mineral oil (64 ton) to suit photo-multiplier tubes (PMTs).   A total of 354 PMTs (Hamamatsu R7081, 10 inch) were mounted in the buffer walls (barrel, top and bottom) pointing inward.   The outermost part is a veto detector (Radius = 4.2 m, Height = 8.8 m) containing purified water (353 ton) and equipped with 64 water-proof PMTs of the same type used in the buffer.   The target, gamma-catcher and buffer are called the inner detector (ID) of RENO and the veto detector is called the outer detector (OD) of RENO.

### 3.2 Energy scale calibration

An energy scale calibration is important in this analysis.   To convert number of photo-electrons (NPEs) collected by PMTs to energy, we used three commercially available radioactive sources with well-known peak energies: $^{68}$Ge (1.022 MeV), $^{60}$Co (2.506 MeV) and $^{252}$Cf (2.2 MeV for Hydrogen capture and 8.0 MeV for Gd capture).   Figure 2 shows the relation between energy (x-axis) of the three radio-active sources and their corresponding NPEs (y-axis) collected in our PMTs.   The four black dots with error bars are data points and the black line is a fit function obtained from the four data points.    The bottom panel shows an accuracy of the fitting and it is within 0.1 % level.    Using the fit functions for the near and far detectors we converted NPEs to energy for all IBD(-like) events.    Figure 3 shows the peak energy value of IBD delayed signal as a function of time, which shows stability of our detector.    More details on calibration will be covered in a separate paper in the near future.

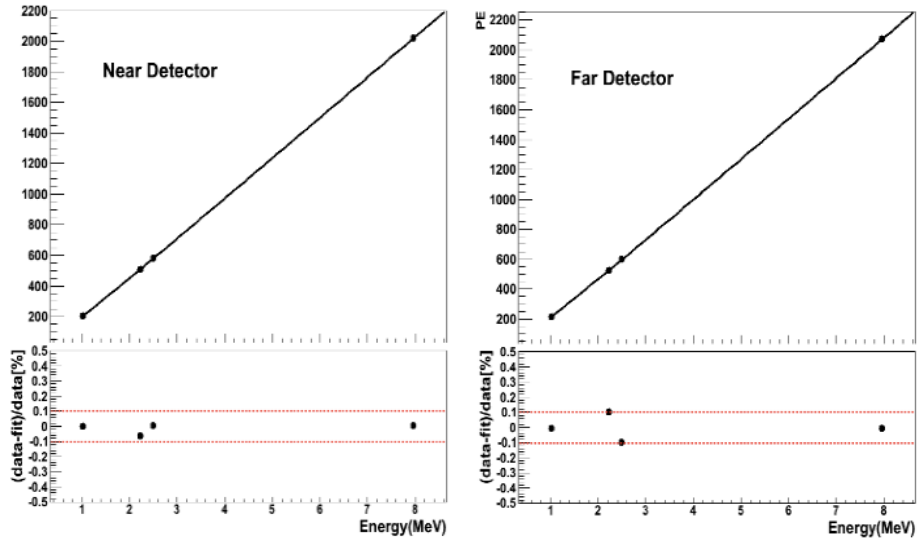

Fig. 2. Top panel: energy of radioactive sources ($^{68}$Ge, $^{252}$Cf: H-capture, $^{60}$Co, $^{252}$Cf: n-capture) in x-axis vs. corresponding NPEs in y-axis.    Black line represents fitting function from the four data points.    Bottom panel: fitting accuracy for each data point is less than 0.1 %.

## 4. Data Analysis

For our new result, we analyzed more data than our first result, slightly less than twice statistics.    We still used rate only analysis method but improved our systematic error.    More details on how we did our analysis are discussed in the following subsections.

### 4.1  Data

RENO has started taking data since August 2011 using both near and far detectors.    We take data almost continuously, and so far (as of Nov. 2013) about 800 live days of data were taken and being analyzed.    For our new result reported in this paper, however, we used 402.7 (369.0) live days of data collected in far (near) detector.    The corresponding data taking period is from August 11$^{th}$, 2011 to October 13$^{th}$ 2012.    The average DAQ efficiency during this period is about 95 %.    In the following subsections characteristics of IBD signal and background, IBD event selection criteria, and systematic error are discussed in a row.

### 4.2  IBD signal and background

As a result of IBD process a positron and a neutron are produced from the interaction of an electron ani-neutrino ( > 1.8 MeV) with a proton target.    Almost all of the energy of the electron anti-neutrino is transferred to positron kinetic energy.    The IBD positron immediately annihilates together with an electron and produces light (1.02 MeV + neutrino kinetic energy) which is registered as a prompt signal (s1) representing reactor neutrino energy.   On the other hand the IBD neutron carrying a few keV from electron anti-neutrinos is thermalized and then captured by Gd with a mean delayed time of 28 usec (in 0.1 % Gd concentration) registering delayed signal (s2) of about 8 MeV.   Thus an IBD signal can be identified by a pair of events, i.e., prompt (s1) and delayed (s2) events, with a mean separation time of 28 usec.

There are three types of backgrounds which mimic IBD signals.   They are accidental, fast neutron and $^9$Li/$^8$He backgrounds.   The accidental backgrounds are caused by external gammas such as radioactivity gamma (from our detector and its environment) mimicking prompt signal, and thermal neutrons (or fast neutrons induced by atmospheric muons) mimicking delayed signal when captured by Gd (or recoiling off protons).   This uncorrelated pair follows a   poisson statistics, and thus accidental backgrounds were estimated using the following relation:

$$N_{accidental} = N_{s2} \times \left(1 - \exp^{[-R_{s1}(Hz) \times \Delta T(s)]}\right) \pm \frac{N_{accidental}}{\sqrt{N_{s2}}}$$

where, s1 and s2 events are counted just before IBD pairing (but after muon removal) by only requiring their corresponding energy range, i.e., [0.75, 12] MeV for s1 and [6, 12] MeV for s2, $R_{s1}$(Hz) is s1 event rate in Hz, and $\Delta T$(s) is a maximum coincidence time window in second, i.e., 1e$^{-4}$ sec .   The estimated accidental background rate in this analysis is 3.61 ± 0.05 /day (0.60 ± 0.03 /day) for the near (far) detector.

Fast neutrons are produced when atmospheric muons pass through rocks surrounding our detector and detector itself.   These fast neutrons recoil off protons resulting in mimicking prompt signal and then are captured by Gd faking delayed signal.   Fast neutrons were estimated in an IBD signal search process with s1 energy spectrum extended up to 30 MeV. Above 12 MeV the s1 spectrum caused by fast neutrons appears flat and thus a fitting with a flat function was performed in [12, 30] MeV region to estimate fast neutron background in the signal region where the same flat spectrum is assumed.   The estimated fast neutron background rate in this analysis is 3.14 ± 0.09 /day (0.68 ± 0.04 /day) for the near (far) detector.

When atmospheric muons entering our detector smashes $^{12}$C in liquid scintillator, $^9$Li and $^8$He are produced.   $^9$Li and $^8$He are unstable isotopes and thus produce (β, n) followers which ends up faking IBD signals.   $^9$Li and $^8$He were estimated using, so called, a "scaling" method.   The concept of the "scaling" method is to estimate $^9$Li/$^8$He background in IBD candidate data using a pure $^9$Li/$^8$He background sample by scaling $^9$Li/$^8$He background above 8 MeV (in IBD candidate data where there is almost no IBD signal events) to that in a pure $^9$Li/$^8$He background sample.    The pure $^9$Li/$^8$He background sample was obtained by selecting events ( > 1.3 GeV) whose time distance with last muon is greater than 500 msec and then subtracting IBD signal portion selected by the same selection criteria.    The estimated $^9$Li/$^8$He background in this analysis is 13.73 ± 2.13 /day (3.61 ± 0.60 /day) for the near (far) detector.

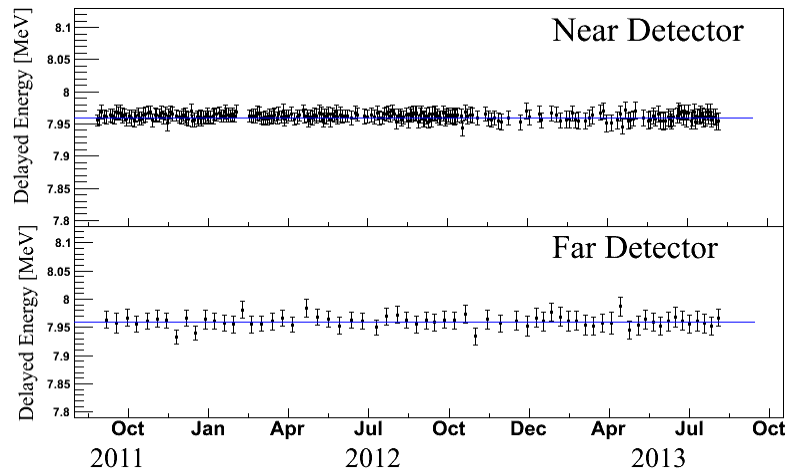

Fig. 3 Time stability of the peak energy (7.96 MeV) of the IBD delayed (s2) signal for the near (top) and far (bottom) detectors.

By combining these three background types the total background rate is estimated as 20.48 ± 2.13 /day (4.89 ± 0.60 /day) for the near (far) detector.    Figure 4 shows energy spectrums of accidental, fast neutron and $^9$Li/$^8$He backgrounds.

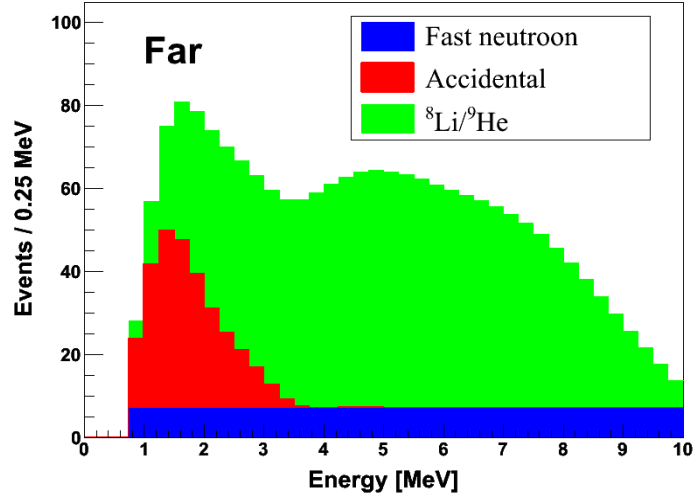

Fig. 4 Stacked histograms of energy spectrums of the three types of background events from data in the far detector.

**4.3 Event selection**

The selection process of IBD candidate events was optimized to keep more signal events while suppressing background events.    The following cuts were applied in sequence: (a) $Q_{max}/Q_{tot}$ < 0.03, where $Q_{max}$ is the maximum charge deposited in a PMT among all PMTs. This cut was devised to eliminate external gammas, i.e., to reduce accidental background;    (b) ("r" >= 0.25 or    $n_{max}/n_{tot}$ <= 0.05) and ("r" <= 0.26 or    $n_{max}/n_{tot}$ <= 0.06) where "r" is called "anti-isolation" cut variable and was devised to remove events from flashing PMTs.    The "r" is defined as a ratio of average NPEs in a neighboring PMT cluster (5~6 PMTs) and maximum charge ($n_{max}$) registered in a PMT in a [-400, 800] μs time window.    The $n_{tot}$ is total charge collected in all PMTs in [-400, 800] μs time window; (c) a cut rejecting events that occur within a 1 ms time window following a cosmic muon traversing the ID with deposited energy ($E_\mu$) larger than 70 MeV, or with    20 MeV < $E_\mu$   < 70 MeV accompanying NHIT > 50 in veto region (OD); (d) events are rejected if they are within a 10 ms time window following a cosmic muon traversing the ID if $E_\mu$ > 1.5 GeV; (e) 0.7 MeV < $E_p$ < 12.0 MeV; (f) 6.0 MeV < $E_d$ < 12.0 MeV where $E_p$ ($E_d$) is the energy of prompt (delayed) signal; (g) 2 μs < $\Delta t_{e+n}$ < 100 μs where $\Delta t_{e+n}$  is the time separation between the prompt and delayed signals; (h) removing any s1 candidate if there is any preceding ID or OD trigger within a 100 μs time window before a prompt event; (i) removing any single (s1 or s2) events if there is any trigger before (300 μs) or after (1000 μs) the single events. This cut is useful to remove fast neutron background.; (j) IBD multiplicity cut rejecting any current IBD pair if there is a following IBD pair within 500 μs from the s2 of a current IBD pair.

After applying these cuts a total of 279,787 (30,211) IBD candidate events with $E_{s1}$ < 10 MeV were selected from the near (far) detector. After subtracting background average daily observed IBD rates become 737.69 ± 2.57 /day and 70.13 ± 0.74 /day for the near and far detectors, respectively, for the period of data we used in this new result. Figure 5 shows our daily observed IBD rate for each day of data taking.

The total detection efficiencies are estimated as 61.99 ± 1.40 % and 71.37 ± 1.19 % for the near and the far detectors, respectively.

**4.4 Systematic errors**

The systematic errors of RENO are estimated from three parts: reactor, detector and background. For the reactors the total systematic errors in our new result remain the same as our first result [12] and they are 2.0 % and 0.9 % for correlated and uncorrelated errors, respectively.

For detectors the total correlated error has slightly improved from 1.5 % to 1.3 % mainly due to the improvement in our "spill-in" systematic error (from 1.0 % to 0.7 %) which was the largest systematic error component in our detector systematic part. The total uncorrelated systematic error of our detectors remain the same as our first result, i.e, 0.2 % which is very small, implying how identical our two detectors are.

For background our total systematic error has improved from 27.3 (17.4) % to 10.4 (12.3) % for near (far) detector as described in 4.2. The most contributor to our background systematic error is $^9Li/^8He$ background and it has improved from 47.6 (29.0) % to 15.5 (16.6) % for near (far) detector. These improvements is due to the new method to estimate $^9Li/^8He$ background described in section 4.2. We still have a room to improve further the $^9Li/^8He$ background systematic error since the error was estimated rather conservatively.

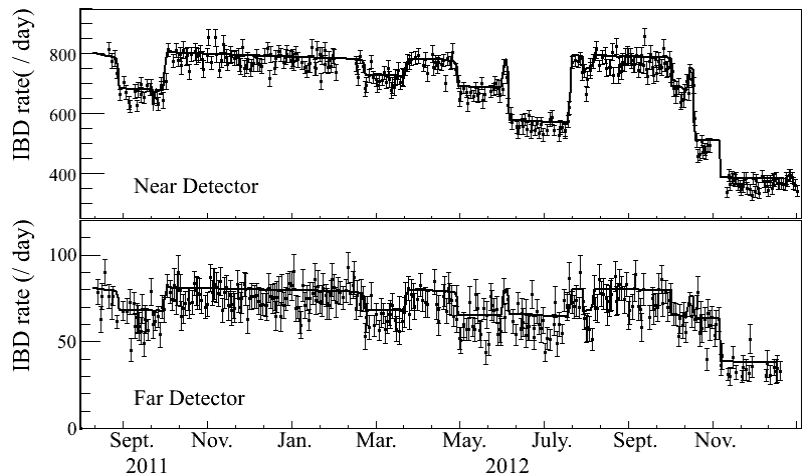

Fig.5 Expected reactor neutrino rate without oscillation (solid line) and observed neutrino rate after background subtraction (dots with error bars). Each dent corresponds to a period when one or more reactors are off.

## 5. Results

Table 1 summarizes our new result: total IBD event rate, background rate, live time and detection efficiency for each detector.   The IBD event rate (after background subtraction) in the near detector was used to estimate the IBD events without oscillation in the far detector (shown as a solid histogram in the upper panel of Fig. 6).   The observed IBD events (after background subtraction) in the far detector is overlaid in the upper panel of Fig. 6 marked as black dots with error bars.   There is a clear deficit in the observed IBD events in the far detector.   This deficit is quantified as R (observed IBD events divided by expected IBD events without oscillation in the far detector) and our measured value is:

$$R = 0.929 \pm 0.006 \text{ (stat.)} \pm 0.007 \text{ (syst.)}.$$

The bottom panel of Fig. 6 shows the ratio R as a function of energy, obtained directly from the ratio of the two histogram in the upper panel of Fig. 6.   The $\sin^2 2\theta_{13}$ value was determined using rate-only analysis with $\chi^2$ function minimization method using pull terms described in [12] and the obtained value is:

$$\sin^2 2\theta_{13} = 0.100 \pm 0.010 \text{ (stat.)} \pm 0.015 \text{ (syst.)}.$$

This is a 6.3 σ significance result against no oscillation hypothesis.

Table 1. Event rates of the observed IBD candidates and the estimated background

| Detector | Near | Far |
| --- | --- | --- |
| Selected events | 279,787 | 30,211 |
| Total background rate (per day) | 20.48±2.13 | 4.89± 0.60 |
| IBD rate after background subtraction (per day) | 737.69±2.57 | 70.13±0.74 |
| DAQ live time (days) | 369.03 | 402.69 |
| Detection efficiency (e) | 0.619±0.014 | 0.714±0.012 |
| Accidental rate (per day) | 3.61±0.05 | 0.60±0.03 |
| $^9$Li/$^8$He rate (per day) | 13.73±2.13 | 3.61±0.60 |
| Fast neutron rate (per day) | 3.13±0.09 | 0.68±0.04 |

## 6. Conclusion and Prospects

Our new result on $\sin^2 2\theta_{13}$ value is consistent with our first measurement but both statistical and systematic errors are improved.   The precision of our new $\sin^2 2\theta_{13}$ measurement is 18 %.   As shortly mentioned in Introduction, precision measurement of neutrino mixing parameters are important.   Therefore we will continue to take data and at the same time to put more efforts to improve our systematic error which is larger than our statistical error. Our final goal is to push our precision on $\sin^2 2\theta_{13}$ value down to 7 % level using 5 live years of data.

We are currently working on analysing about 800 live days of data using both shape and rate analysis methods. Other analyses such as absolute neutrino flux measurement and Hydrogen-captured IBD analyses are also underway.

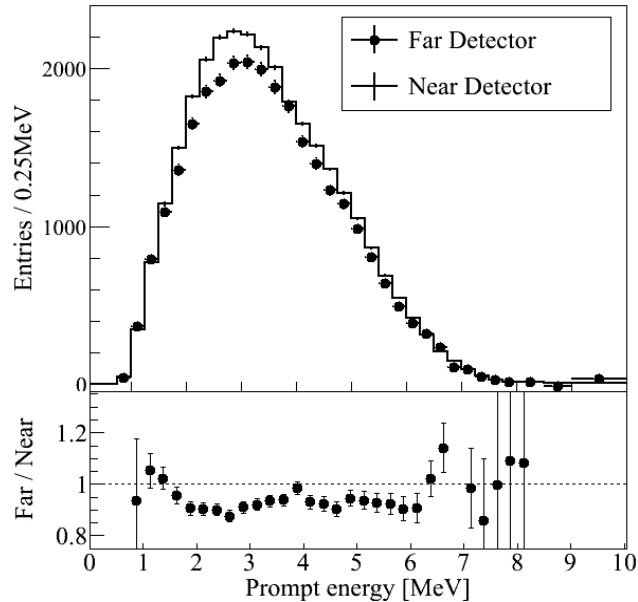

Fig. 6  Top panel: observed energy spectrum of the prompt (s1) signal events in the far detector (dots) compared with expected s1 spectrum with no oscillation at the far detector estimated using neutrinos observed in the near detector (histogram). Errors are statistical uncertainties only. Bottom panel: The ratio of the measured spectrum of the far detector to the non-oscillation prediction, i.e., ratio of the two histograms in the top panel.

## Acknowlegement

This work was supported by the National Research Foundation of Korea (NRF) grant funded by the Korea government (Ministry of Science, ICT & Future Planning) (No. 2009-0083526).